%% file: cas-dc-template.tex
\documentclass[a4paper,fleqn]{cas-dc}

\usepackage[numbers]{natbib}

\usepackage{caption}
\usepackage[textwidth=  15mm]{todonotes}
\usepackage{float}
\usepackage{hyperref}
\usepackage{xcolor}
\usepackage{soul}
\usepackage{rotating}
\usepackage[ruled,vlined]{algorithm2e} 
\graphicspath{{./images/}}

\begin{document}
\let\WriteBookmarks\relax
\def\floatpagepagefraction{1}
\def\textpagefraction{.001}

\shorttitle{Differential Area Analysis for Ransomware Detection}

\shortauthors{S.R.Davies, R.Macfarlane, W.J.Buchanan}

\title [mode = title]{Differential Area Analysis for Ransomware Attack Detection within Mixed File Datasets}    

\author[1]{Simon R. Davies}[type=editor,
                        auid=000,bioid=1,
                        orcid=0000-0001-9377-4539]

\cormark[1]
\ead{s.davies@napier.ac.uk}
\address[1]{School of Computing, Edinburgh Napier University, Edinburgh, UK}

\author[1]{Richard Macfarlane}[]
\author[1]{William J. Buchanan}[auid=002,bioid=2,
                        orcid=0000-0003-0809-3523]


\input{01abstract}

\begin{keywords}
Entropy \sep Ransomware Detection \sep Test Data Sets 
\sep
\sep
\sep
----------------------------------------------
\sep
Article history\
\sep
\begin{tabbing}
\hspace{24mm}     \= \hspace{1mm} \= \kill
\end{tabbing}
\end{keywords}

\maketitle
\input{02introduction}

\input{03literaturereview}

\input{04design}

\input{05implementation}

\input{06evaluation}
\input{07conclusions}

\printcredits


\bibliographystyle{IEEEtranS}

\bibliography{bibliography}

\end{document}

%% file: 01abstract.tex
\begin{abstract}
\newline
\noindent The threat from ransomware continues to grow both in the number of affected victims as well as the cost incurred by the people and organisations impacted in a successful attack.
In the majority of cases, once a victim has been attacked there remain only two courses of action open to them; either pay the ransom or lose their data. One common behaviour shared between all crypto ransomware strains is that at some point during their execution they will attempt to encrypt the users' files. This paper demonstrates a technique that can identify when these encrypted files are being generated and is independent of the strain of the ransomware.

An enhanced mixed file ransomware data set of more than 130,000 files was developed based on the govdocs1\citep{Garfinkel} corpus. This data set was enriched to contain examples of files that reflect the more modern Microsoft file formats, as well as examples of high entropy file formats such as compressed files and archives. The data set also contained eight different sets of files that were generated as the result of different real-world high profile ransomware attacks such as WannaCry, Ryuk, Phobos, Sodinokibi and NetWalker.

Previous research~\cite{Penrose2013,Zhao2011} has highlighted the difficulty in differentiating between compressed and encrypted files using Shannon entropy as both file types exhibit similar values. One of the experiments described in this paper shows a unique characteristic for the Shannon entropy of encrypted file header fragments. This characteristic was used to differentiate between encrypted files and other high entropy files such as archives. This discovery was leveraged in the development of a file classification model that used the differential area between the entropy curve of a file under analysis and one generated from random data. When comparing the entropy plot values of a file under analysis against one generated by a file containing purely random numbers, the greater the correlation of the plots is, the higher the confidence that the file under analysis contains encrypted data. The experiments demonstrate a high degree of confidence in the accuracy of the model achieving a success rate of more than 99.96\% when examining only the first 192 bytes of a file, using a mixed data set of more than 80,000 files. This technique successfully addresses the problem of using file entropy to differentiate compressed and archived files from files encrypted by ransomware in a timely manner. 
\end{abstract}

%% file: 02introduction.tex
\section{Introduction}
\label{cha:intro}

\noindent Ransomware is a malicious class of software that utilises encryption as a method  to implement an attack on system availability. The victim's data remaining encrypted and held captive by the attacker until the ransom demand is met~\cite{Bajpai}.

The occurrence of ransomware infection continues to\\  grow in scale, cost, complexity and impact since its initial discovery nearly 30 years ago~\cite{Young1996,Young2017}. Security practitioners are engaged in a continual arms race with ransomware developers attempting to defend their digital infrastructure against such attacks. 
During the recent changes in working practices due to COVID-19 where an increasing number of employees are working from home, ransomware attacks have increased, and attack vectors have changed. Currently the most common attack vectors are Remote Desktop Protocol~(RDP) and phishing emails~\cite{Coveware}, and average ransomware payments have increased by over 60\% in the second quarter of 2020 to now be in excess of \$170,000 per incident.
Examples of recent high profiles attacks where it is believed that the ransom has been paid are the Wastedlocker strain that attacked Garmin~\cite{Porter2020}, the Netwalker strain that attacked the University of Utah~\cite{ODonnell2020b} and the Sodinokibi/Revil strain that attacked the owner of the Jack Daniel's distillery~\cite{Seals2020}.

Ransomware infection detection tends to fall into one of these approaches:
\begin{itemize}
    \item Behavioural/Dynamic analysis. This is where the suspicious processes behaviour is monitored to see if it follows the profile of known attacks. Many techniques have been proposed using this approach for detection and infection prevention. These can be classified into four broad categories: file-based detection, system-based behaviour detection, resource-based behaviour detection, and connection-based behaviour detection \cite{Al-rimy2018, Alekseev2017, Egele2012, McIntosh2018}. Candidates for monitoring could be CPU usage, disk access, system calls and attempted external communications.
    \item Signature/Static analysis. This is where the executable code is analysed prior to its execution in an attempt to identify known sequences of bytes that indicate the probability that the program is malicious. 
\end{itemize} 

Obfuscation techniques and polymorphism~\cite{Al-rimy2018,McIntosh2018} have been used by ransomware for some time in an attempt to avoid signature detection. With regards to the dynamic detection approach, the behaviour of different ransomware stra-ins varies significantly making them difficult to profile and this approach is often ineffective against unknown strains. While some strains are self-contained - knowing which tasks to perform and having the encryption keys included in the binary - others attempt to contact external command and control (C\&C) servers for guidance on how to proceed. Some strains also exfiltrate data or attempt lateral movement prior to encryption. What is becoming increasingly obvious is that as the arms race continues and new strains incorporate an increasing amount of these diverse techniques into each iteration, thus the difficulty in ransomware infection detection continues to be a problem.

Many ransomware detection methods have a low detection rate. They also suffer from having high false positive rates, where they flag benign programs as malicious, and false-negative rates, and thus fail to identify malicious programs. Also, as current detection techniques need to collect significant amounts of information while monitoring, they have the disadvantage that they can consume a large amount of system resources~\cite{Lee2019}. 

One thing that is common amongst all crypto-ransomware strains is that, at some point, they will attempt to encrypt the users' files and write these encrypted files to disk. It is this common feature of ransomware execution that the experiments in this paper investigate. Continella et al.~\cite{Continella} define that filesystem access is a strategic point for monitoring typical ransomware activity.

This paper demonstrates a technique that could be used to rapidly identify the creation of these encrypted files. The proposed technique tests only the first few bytes of the file being written and performs analysis on this file sample to determine if the file being written is encrypted or not. Similar to other research~\cite{Alekseev2017,Continella,Hall2006,Jung2018,Lee2019,Scaife2016,Zhao2011}, initially this analysis will focus on the Shannon entropy~\cite{Shannon1948}, but unlike previous research it will only be performed on the file's header. The paper describes a differential area analysis technique which compares the entropy values of the file under analysis against the entropy values of reference file that contains random data.
It is envisaged that this technique could be expanded to include other types of tests along with Shannon entropy. The described technique could in theory, be used to alert the user of suspicious behaviour, prevent the files from being written or trigger some further live forensic investigation~\cite{Davies2020}. While this technique would not prevent any data exfiltration prior to the start of encryption, it may well be useful in preventing the actual encryption of the user's data, and thus mitigating a significant portion of the attack.

\subsection{Paper Contribution}
\noindent The use of file entropy  as a reliable method for encrypted file identification has been previously investigated by several researchers~\cite{Alekseev2017,Continella,Hall2006,Lee2019,Scaife2016,Zhao2011}. When evaluating the proposed detection methods in the reviewed research, it appears that only a limited subset of file types was used and no examples were found where compressed files formed part of the test data set. The proposed techniques also tended to focus on the overall entropy of the file with only Jung and Won~\cite{Jung2018} examining the entropy of PDF file fragments.

It is a relatively straightforward task, though resource-intensive, to use a file's overall entropy value to differentiate an encrypted file with the majority of the most commonly used file types. However, using the files entropy value to differentiation between compressed files and encrypted files becomes problematic as the overall entropy values are normally similar for these file types.
This paper thus focuses on this specific problem of entropy similarity and proposes a technique that will allow the rapid identification of encrypted files.
This ability to be able to reliably detect encrypted file creation can then be used as an input for a ransomware detection technique.
The main contribution of this paper is to define a classification model that can be used to reliably classify encrypted data files, even amongst a data set that contains files with a high overall entropy.

\subsection{Paper Structure}
\noindent The remainder of the paper is structured as follows. Section 2 contains a review of related work. Section 3 describes the design philosophy followed in Section 4 with a description of the experiment implementation. A critical analysis of the experimental results and comparison to similar work in the field is provided in Section 5. The paper closes in Section 6 with a discussion of the findings and suggestions for further research.

%% file: 03literaturereview.tex
\section{Related Work}

\noindent Previous uses of entropy have focused on processing the file as a whole. With this approach, a particular file (or file type) can be characterised by a bit value representing its information content. For example, text files containing written English have been identified as having a file entropy in the range of 3.25 to 4.5 bits. Compressed files, such as ZIP archives, have a higher entropy level, typically just over six bits~\cite{Hall2006}. In some cases in malware detection, the entropy of the entire file being written has been used as an indicator for the presence of malicious activity. Products such as DropIT~\cite{Scaife2016}, ShieldFS~\cite{Continella} and  Unveil~\cite{Kharraz2016} check the entropy of the whole file being created, and combined with other variables such as the file's magic number and file extension, decide on whether to allow the file to be created. However, the use of file extensions as an indicator is open to abuse as an adversary  can easily change the extension of a file or its magic number at any time, preventing the operating system from identifying the file and allowing the attacker to circumvent the detection~\cite{McDaniel2003}. 
A major issue when using entropy for file type classification is raised by Zhao et al.~\cite{Zhao2011} where they state that when considering entropy as a gauge, most compressed and encrypted data share similar characteristics and they confirm that more work is required to investigate the application of entropy to these file types.

\subsection{File fragment analysis}
\noindent A similar, but related field of research, is in file fragment classification. This involves research into determining the type of a target file by only analysing a portion or a fragment of the file. This is particularly useful in the fields of digital forensics. Some methods used are entropy, n-gram analysis, statistical analysis, machine learning and support vector machines~\cite{Cleary2018}.
McDaniel and Heydari~\cite{McDaniel2003} describe a method of using byte frequency analysis of a file's header and footer to build a \emph{fingerprint} that can be used to identify files of a similar type. During their testing, they identified that it was a viable approach but that it needs to be combined with other methods to improve accuracy.
They also postulate that file header/trailer analysis (FHT) could be used for ransomware executable detection as only a small sample is required. They found  that for successful identification, a fingerprint size of only 53 bytes was required. The algorithm could possibly be of use in cryptanalysis and also to automatically differentiate between real data and encrypted samples~\cite{McDaniel2003}.

With regards to file fragment classification, it has been noted that the structure of some file types, for example, GIF or PPT, results in some areas of the file having lower entropy than other areas within the same file. For example, a file type which has a header containing file metadata in text format with compressed data in the body would differ from a binary file of purely random values. The lower entropy at the start of the file would distinguish it from a file of consistently high entropy. To account for these structured file formats, and to make them distinguishable from other files with similar overall entropy values, a sliding window approach to measurement has been evaluated~\cite{Hall2006}.

Li et al.~\cite{Li2005} used a similar approach where the file types were determined by only analysing either the first 20 or the first 200 bytes of a file using n-gram analysis. They reported that their tests successfully identified files and found that their technique performed as well, if not better, than analysing the whole file, with superior computational performance~\cite{Li2005}. However, no compressed or encrypted files were used as part of their test data. The work performed in this paper differs from Li's work in that instead of using n-gram analysis, we will be using the standard Shannon entropy of the file header as the indicator of encryption, as well as having a large variety of encrypted and compressed files in the data set. It was considered to use an improved Shannon entropy based on 2~KB blocks to allow differentiation between compressed and encrypted files~\cite{Held2018}, but this was rejected due to the small sample window being used. Jung and Won~\cite{Jung2018} did perform analysis of entropy on file fragments including file header and trailer and while their findings were promising, they had limited their investigation to the PDF document format.

\subsection{Randomness, Encryption and Ransomware}
\noindent One of the aims of encryption is to transform useful data into a ciphertext having the property of appearing completely random with no obvious pattern or relationship to the original data~\cite{Aumasson2018}. The more random the ciphertext appears, the better the encryption.

Statistical tests can be employed to determine the randomness of a file, with higher randomness suggesting that the contents may be encrypted. Creation of encrypted files could be the consequence of the action of a ransomware attack so detecting the presence or creation of these files could be used as part of a ransomware detection technique. One drawback with this approach is that the content of some legitimate files such as archive and image files can also appear to be very random.

Generally, the problem of detecting ransomware can be reduced to the problem of detecting random data being written to the file system.  Random (or encrypted) data should comprise of an approximately equal number of each byte, unpredictably distributed across the data. 

Therefore, it is possible to apply widely-used tests for randomness across these byte distributions to identify the presence of random data. This may then indicate the presence of encryption, and possibly a ransomware attack~\cite{Pont2020}.
Statistical techniques~\cite{Pont2020} that can be used to identify the randomness of a data stream are shown below:

\begin{itemize}
\item \textbf{Chi-Square} A test that measures how a model compares to an actual distribution. When considering a test for randomness for Chi-square we would expect an equal frequency of byte values for the expected distribution
\item \textbf{Arithmetic Mean} This is the sum of all the individual bytes in the sample divided by the total number of bytes in that sample. As all the possible byte values range between zero and 255 then the closer this calculated value is to 127.5 then in theory the more random the data under investigation is.
\item \textbf{Monte Carlo}This technique is based on the repeated random sampling of the data under investigation and then performing a time averaging statistical analysis on these samples in order to make a decision on the data.
\item \textbf{Serial Byte Correlation Coefficient}~\cite{Pont2019} A lightweight statistical test that looks at the relationship between consecutive numbers.

\item \textbf{Shannon Entropy}\cite{Shannon1948} In information theory, entropy is a measure of a given input's level of uncertainty and is considered to be an indication of the amount of information contained within each byte.
\end{itemize}

%% file: 04design.tex
\section{Design}
\label{cha:design}
\noindent The work in this paper could be considered a special case of file fragment analysis where only one fragment is analysed and is limited to the first part of the file. 
Instead of creating specialised classifications for each file type as proposed by Roussev and Garfinkel~\cite{Roussev2009}, or using a generated \emph{fingerprint}~\cite{McDaniel2003}, the experiments described in this paper will use the Shannon~\cite{Shannon1948} entropy calculation of the file fragment.

Based on the calculated value, the technique will judge if the file is encrypted, where a compressed file will be classed as an unencrypted file. This approach is supported by, amon-gst others, Alekseev and Platonov~\cite{Alekseev2017} who confirm that testing for entropy is a good indicator for identifying encrypted files. 

The aim of the experiments described in this paper is not to accurately identify the specific file type ~\cite{Fitzgerald2012,Hall2006,Lee2019,Lee2019a,Li2005,McDaniel2003} rather it is to identify if the file is encrypted or not.

\subsection{Data Set}
\label{datasets}
\noindent To facilitate the repeatability of the experiments performed in this paper, it was necessary that a well respected, realistic and easily accessible standard data set be used for the experiments~\cite{Garfinkel2009}. This approach is confirmed by~\cite{Garfinkel2010,Roussev2009} who stress that test data must be representative of data likely to be encountered in real-world situations. It is a known issue within the research community~\cite{Al-rimy2018,Grajeda2017,Maigida2019,Pont2019} that there is often a lack of readily available, researchable ransomware data sets.

Garfinkel et al.~\cite{Garfinkel2009} created a publicly available govdocs1 corpus containing one million files collected using random searches of the .gov domain. These files are stored in 1,000 directories, each containing 1,000 files, and also 10 randomly assigned streams for development and testing purposes. This corpus can be obtained from~\cite{Garfinkel}. Many recent digital forensics studies~\cite{Cleary2018,Fitzgerald2012,Grajeda2017,Kolodenker2017,Nguyen2014,Penrose2013,Pont2020,Pont2019} also obtained their test data from Garfinkel et al.'s~\cite{Garfinkel2009} govdocs1 data set, supporting the claim that this data set is a well-known and respected source of test data.

An advantage of this data set is that it does not contain files that change frequently, such as ransomware or virus samples or files that have been encrypted using these samples. The inclusion of these files would cause the data set to quickly become outdated. Even though the govdocs1 data set is well known and respected, there remain some concerns regarding its use, not least the fact that it is now more than 10 years old. While the govdocs1 corpus is a broad well-known mixed file data set containing more than 55 different file types, the number of some specific file types are not significant. For example, there are only a total of 40 files in the newer Microsoft DOCX and XLSX formats.

A class of files known as archives are also poorly represented. Archive files are normally compressed and are often used to collect multiple files together into a single file for easier portability or to use less storage space. Archive files often store directory structures, error detection, and sometimes use built-in encryption. There are multiple types of archives currently in use, with varying properties and characteristics, however, the only type present in the original data set is \emph{gzip}. \\

\noindent To address the identified shortcomings of the govdocs1 mixed file data set, the following enhancements were performed prior to its use in this research:

\begin{itemize}
\setlength\itemsep{1.2em}
    \item All existing DOC and XLS files present in the govdocs1 corpus are converted to have a corresponding instance in the new Microsoft document format DOCX and XLSX.
    \item 
    Several sets of various archive types were created. Each set containing 1,000 examples of the specific archive type and used the govdocs1 corpus as their source files. They were created using different compression programs such as tar, 7zip, WinRAR, WinZip, as well as archives using different compression options such as high compression, header compression and encryption. In most cases the tool's default configuration values were used. For WinZip this was the DEFLATE algorithm with one pass. 7zip used the Lempel–Ziv–Markov chain algorithm(LZMA2) with a dictionary size of 16MB and a standard Branch Converter Filter(BCJ). For high compression the LZMA2 algorithm was also used but with a dictionary size of 64MB and a newer version of the standard Branch Converter Filter(BCJ2), producing the tools maximum available compression.
    
    \item The govdocs1 files were encrypted by different ransomware samples to produce data sets from each of the selected ransomware strains. Once the ransomware had completed executing, the files affected were gathered into a data set for that ransomware sample. The ransomware strains used during this experiment were: BadRabbit, Netwalker, NotPetya, Phobos, Ryuk, Sodinokibi, WannaCry and WastedLocker.
    \item A synthetic data set of 1,000 files was created of files with lengths varying between 512 and 2,048 bytes that contained pseudo-random data generated using the \\ Python \texttt{'os.urandom'} function. When generating random numbers, pseudo-random number generators use mathematical algorithms, whereas true random number generators use unpredictable physical means, for example, atmospheric noise~\cite{random}.
    The purpose of this is to provide a baseline data set that contained pseudo-random data which could be used as a comparison for the encrypted and compressed data files.
\end{itemize}

\subsection{Entropy}
\noindent File entropy refers to a specific measure of randomness. One such measure is called \emph{Shannon Entropy}~\cite{Shannon1948} and is used to express information content. 
This value is essentially a measure of the predictability of any specific byte in the file, based on preceding bytes~\cite{Rosetta}. It is basically a measure of the \emph{randomness} of the data in a file - measured in a scale of zero to eight (eight bits in a byte). Typical text files that contain only alphanumeric characters and no formatting will have a low value, whereas encrypted or compressed files will have a high measure~\cite{VandenBrink2016}.

Another way to consider entropy is that it is a measure of the predictability or randomness of data. A file with a highly predictable structure or a bit pattern that repeats frequently has low entropy. Such files would be considered to have low information content (or low information density). Files in which the next
byte value is relatively independent of the previous byte value would be considered to have high entropy. Such files would be thought to have high information content~\cite{Hall2006,Schneier1996}.

The maximum possible entropy per byte is eight bits of entropy per byte signifying that it is completely random. 
The generally accepted formula for entropy ($H$)~\cite{Shannon1948}, is given as:
\begin{equation}
H(\textit{X})=-\sum_{i=1}^{n} P(x_i)\,log_2\,P(x_i)
\label{eq:Shannon}
\end{equation}
where $H$ is the entropy (measured in bits), $n$ is the number of bytes in the sample and $P(x_i)$ is the probability of byte \emph{i} appearing in the stream of bytes. The negative sign ensures that the result is always positive or zero.

\subsection{Experimental design}
\noindent The experiments described in this paper can be broken down into two distinct areas. Initially, to discover the overall entropy profiles for different file types, the method analyses only the first sequence of bytes of the files under investigation. Once these entropy values have been determined, the second experiment aims to correctly identify encrypted files in a data set.

\subsubsection{Entropy profile}
\begin{figure}
    \centering
    \includegraphics[width=0.5\textwidth]{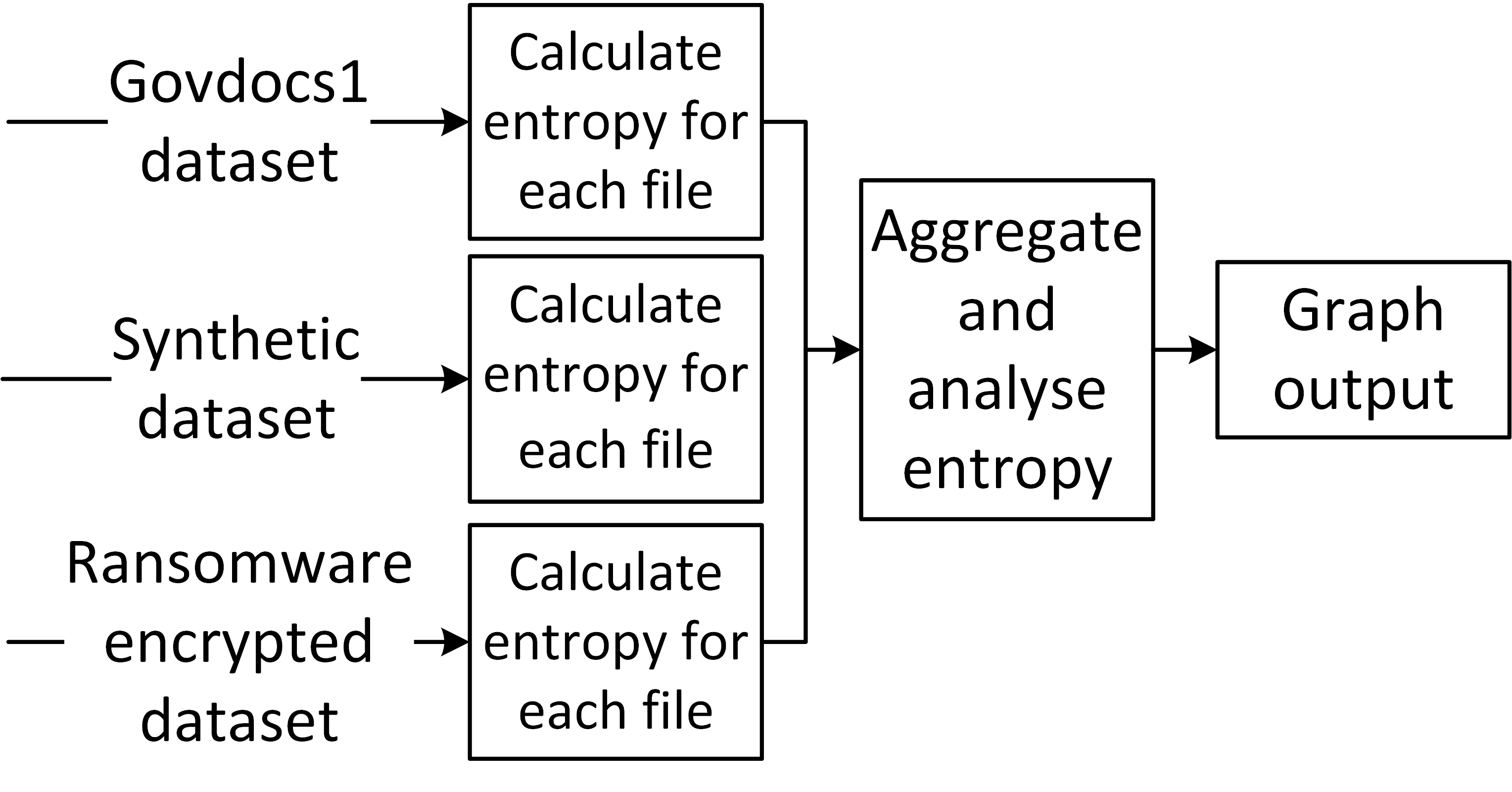}
    \caption{Experiment Overview.}
    \label{fig:Oveview}
\end{figure}

\noindent The header of each file found in the test data set is processed in turn. Each selected file is analysed 32 times, starting with the first eight bytes of the file header, the first 16, the first 24, and so on, up to the first 256 bytes in eight-byte increments. For each analysis, the entropy of the sampled bytes was calculated and recorded. The averages of the entropy calculations are then grouped by byte length and file type. A graphical representation of this experiment is shown in Figure~\ref{fig:Oveview}.

A basic outline of the experimental steps performed is illustrated in Algorithm~\ref{alg:algo1}.

\begin{algorithm}
\SetKwInOut{Input}{Input}
\SetKwInOut{Output}{Output}
function \underline{EntropyPlot} $(files)$\;
    \Input{List of files $(files)$ to analyse}
    \Output{list of arrays contain x,y plot coordinates}

plot\_coord = ();\\
 \ForEach{file\_type of files}{%
 
  \ForEach{f of  file\_type}{%
  buff = open(f)\;
  \For{$len=8;$len<=256;len=len+8}{
    fragment=buff[len] ;\\
    \tcp{entropy of fragment}
    file\_plot[len] = $log_2(255 \times fragment)$;\\
  }
  close(f);
  }
  \For{$len=8;$len<=256;len=len+8}{
    plot\_coord(file\_type)[len] =  
        average( file\_plot[len]);\\ 
    }
 file\_plot=[];\\
 }
return plot\_coord;
 
 \caption{Header Entropy Test Procedure}
 \label{alg:algo1}
\end{algorithm}

\subsubsection{File classification}
\noindent In the second experiment, each file is analysed to determine how closely its entropy values match a file that contained purely random numbers. A classification decision is then made based on the similarity of these two.

\begin{figure}
    \centering
    \includegraphics[width=0.45\textwidth]{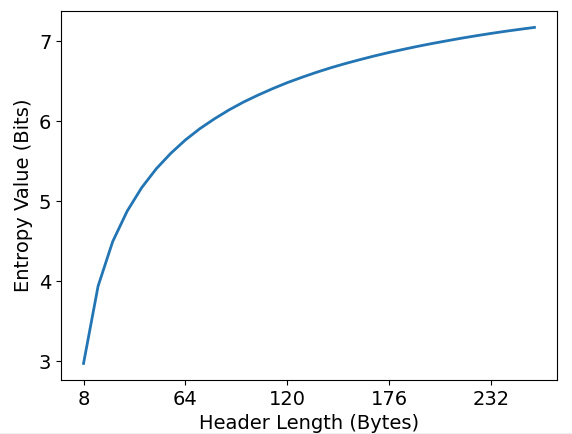}
    \caption{Entropy value plot for a file containing purely random data.}
    \label{fig:random}
\end{figure}

When applied to file fragments, the resulting calculated Shannon entropy value for small sample sizes is typically relatively low and increases as the sample size increases. This is illustrated in Figure~\ref{fig:random} which shows a plot of the Shannon entropy values for different sample sizes taken from a file that contains random numbers. The byte length of the file fragment analysed is shown in bytes along the x-axis. This plot represents an ideal curve for random data, but it may be closely approximated to encrypted data. As previously stated, encrypted data share similar Shannon entropy ranges to random data. 

One method that can be used to quantify the closeness to random samples would be to compare it to a plot generated by a reference file that contains purely random numbers. While there are several possible techniques available to analyse the variations, such as the Chi-square~\cite{Pearson1900} method, the technique selected for these experiments determines the differential area between the sampled target file and a reference sample of random data. 
This approach is selected over just calculating the difference between the value at each point, as the area calculation takes into account how close the whole curve matches over the complete interval and not just at a specific sampling interval.

\begin{figure}
    \centering
    \includegraphics[width=0.45\textwidth]{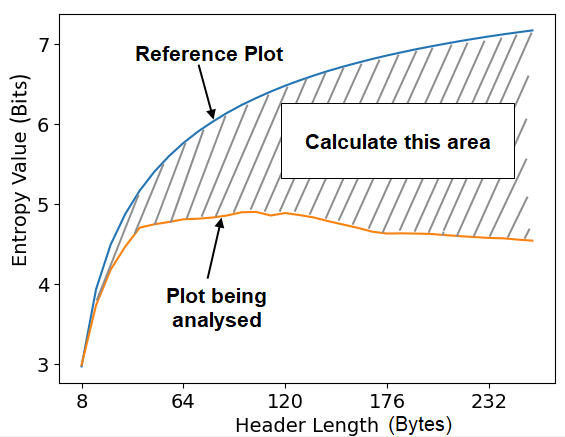}
    \caption{Plot differences.}
    \label{fig:plotdiffernces}
\end{figure}

The values on the y-axis are given in bits and the values on the x-axis are given in bytes, resulting in the area unit being Bit-Bytes. This calculated \emph{Bit-Byte} area value is then used to determine if the file represented by the entropy target file contains encrypted data or not. The lower the area value, the closer the sampled target file matches the randomised data source, the more likely it is that the file contains random or encrypted data. A graphical representation of this analysis is shown in Figure~\ref{fig:plotdiffernces}.

\begin{figure}
    \centering
    \includegraphics[width=0.45\textwidth]{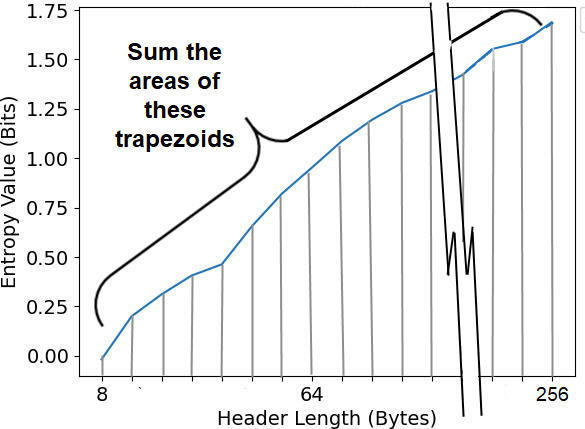}
    \caption{Trapezoidal area of the derived plot.}
    \label{fig:traparea}
\end{figure}

This area calculation can be achieved in reality by generating a new third plot. The values of which are derived by subtracting the plot under analysis values from the random data plot values. The area between the x-axis and this new third plot can then be calculated. A graphical representation of this calculation is shown in Figure~\ref{fig:traparea}.

The newly created plot is not a precise curve, rather it is a series of straight lines joining each sample point. Due to this property and using an equal step size, the optimum method for calculating the area under the plot is to use the \emph{Composite Trapezoidal Rule}~\cite{Atkinson1989} given as: 

\begin{equation}
  I_{composite}=\frac{h}{2}\left[f(a)+2 \sum f(x+h_i)+f(b)\right]
  \label{eq:TrapRule}
\end{equation}
where $a$ is the start point, $b$ is the endpoint and $h$ is the step value between the plot points.

The closer the plot under analysis is to the reference plot, the lower the calculated area value will be and the higher the confidence is that the file contains random or encrypted data.
A basic outline of the experimental steps performed is illustrated in Algorithm~\ref{alg:algo2}.
\begin{algorithm}
\SetKwInOut{Input}{Input}
\SetKwInOut{Output}{Output}
function \underline{FileClassification} $(files)$\;
    \Input{List of files $(files)$ to analyse}
    \Output{list of pass/fail statistics for each header length an file type}

\tcp{Shannon values for random data plot}
rand\_plot = [x y] \;
stats = ();\\
 \ForEach{threshold=start;threshold < max; threshold++}{
 \ForEach{file\_type of files}{%
 
  \ForEach{f of  file\_type}{%
  buff = open(f)\;
  \For{len=8;len<=256;len=len+8}{
    fragment= buf[len]\;
    \tcp{entropy of fragment}
    e = $log_2(255 \times fragment)$\;
    \tcp{subtract from random plot}
    plot[len] = rand\_plot[len] - e\;
  }
  close(f)\;
  area = area(plot);\\
  \eIf {area<=threshold}{
    stats(file\_type)(threshold)(pass) +=1 \;
    }{
    stats(file\_type)(threshold)(fail) +=1\;
    }
  }
 }
 }
 return stats;
 \caption{File Classification}
 \label{alg:algo2}
\end{algorithm}

When classifying the file types there are four possible outcomes. Possible classifications are shown in Table~\ref{tab:classification-options}. The proposed model uses these classification values to determine the overall accuracy of the model~\cite{Ting2017}. The formula to compute the \ul{ Accuracy}($ACC$) is given as:
\begin{equation}
  \textit{ACC} = \frac{TP + TN}{TP+TN+FP+FN}
    \label{eq:Accuracy}
\end{equation}

\begin{table}
\caption{Possible classification outcomes}
\small{
\begin{tabular}{l@{}|l}
\hline
 \textbf{Classification} & \textbf{Description} \\
\hline
True Positive\hspace{2.4mm}(TP) & Correctly classified  encrypted file \\
True Negative\hspace{1.5mm}(TN) & Correctly classified non-encrypted file \\
False Positive\hspace{2mm}(FP) & Classified normal file as an encrypted file \\
False Negative\hspace{1mm}(FN) & Failed to classified encrypted file \\
\hline
\end{tabular}
}
\label{tab:classification-options}
\end{table}

\subsubsection{File classification example}
\noindent To illustrate the proposed method in more detail, the following example will be used. In this example, we will be analysing entropy plots for header lengths of 40 bytes and assuming that the variable threshold is 20 Bit-Bytes. If the calculated area is below the threshold value, it is assessed to be an encrypted file. As a sample for analysis, entropy values are taken from a file generated using the 7Zip program, new plot points for a derived graph are calculated by subtracting the values for the graph under analysis from the points that describe a file that contains purely random data. The values and the results from this analysis is shown in Table~\ref{tab:example-plot-points}.

\begin{table}
\caption{Calculation of the derived plot entropy values}
\small{
\begin{tabular}{l|l @{\hspace{1.5\tabcolsep}} l @{\hspace{1.5\tabcolsep}} l @{\hspace{1.5\tabcolsep}} l @{\hspace{1.5\tabcolsep}} l @{\hspace{1.5\tabcolsep}} l}
\hline
 Header length & 8 & 16 & 24 & 32 & 40 \\
 \hline
Random plot (y)& 2.976 & 3.941 & 4.496 & 4.878 & 5.171 &   \\
Sample plot (y) & 3.000 & 3.807 & 3.336 & 3.430 & 3.903 &   \\
Derived plot (y) & -0.024 & 0.134 & 1.160 & 1.448 & 1.268 &  \\
\hline
\vspace{5 mm} 
\end{tabular}
}
\label{tab:example-plot-points}
\end{table}

The sum of the areas of each trapezoid of the new plot is then determined using equation~\ref{eq:TrapRule} where, \emph{a=8}, \emph{b=40} and \emph{h=8}. For clarity, the individual trapezoidal areas are shown in Table~\ref{tab:example-plot-area}.

\begin{table}
\caption{Calculated area for each trapezoid}
\begin{tabular}{l|lllll}
\hline
 Header length & 16 & 24 & 32 & 40 \\
 \hline
Trapezoid Area & 0.44 & 5.176 & 10.432 & 10.864 \\
\hline
\end{tabular}
\label{tab:example-plot-area}
\vspace{5 mm} 
\end{table}

The calculated area in this example gives a final value of \textbf{26.91} Bit-Bytes.
This value is then compared to the threshold value and a classification decision is made.
In this case, the value is greater than the threshold, so this file is correctly classified as a non-encrypted file, also known as True-Negative (TN) classification. If the area had been below the threshold of 20 Bit-Bytes, it would have been incorrectly classified as an encrypted file and would have been a False-Positive (FP) classification.

%% file: 05implementation.tex
\section{Implementation}
\label{cha:implementation}

\noindent A subset of files taken from the first 100 archives of the complete govdocs1~\cite{Garfinkel2009} archives were initially selected as the input for these experiments together with the additional data sets for XLSX/DOCX, compressed files and ransomware encrypted files mentioned in Section \ref{datasets}.
When generating the data sets relating to specific ransomware samples, a set of govdocs1 files were placed on a completely isolated target machine and the ransomware sample executed, paying particular attention to the ethical consequences of running such programs. Once the ransomware program execution had completed, the files placed on the machine were examined to determine if they had been updated by the ransomware's execution. If they had been changed, they were then added to the data set for that ransomware sample. Different ransomware strains can target different file types, which can result in different data set sizes for different ransomware strains.

Programs written in Python were created to perform the experiment steps outlined in algorithms~\ref{alg:algo1} and ~\ref{alg:algo2}. The entropy calculation algorithm used was adapted from the code developed for the \texttt{'findaes'} ~\cite{Kornblum} and \texttt{'interrogate'} ~\cite{Maartmann-Moe2009} programs, and are based on the work by Trenholme~\cite{Trenholme2014}. Some additional validation of the results was also performed by running the same calculations via the website~\cite{Asecuritysite}.

\begin{table}
\caption{Excluded file entropy values for the first 40 bytes.}
\renewcommand{\arraystretch}{1.2}
\scalebox{0.90}{
\begin{tabular}{ l | r | c | l | r | c }
\hline
\textbf{\begin{tabular}[c]{@{}l@{}}File \\ Type\end{tabular}} & \textbf{\begin{tabular}[c]{@{}l@{}}Sample \\ Size\end{tabular}} & \textbf{Entropy} & \textbf{\begin{tabular}[c]{@{}l@{}}File \\ Type\end{tabular}} & \textbf{\begin{tabular}[c]{@{}l@{}}Sample \\ Size\end{tabular}}  & \textbf{Entropy} \\
\hline
csv & 1,032 & 3.849  & ppt & 5,642 & 2.267 \\
data & 2 & 2.374     & sgml & 9 & 3.765 \\
dbase3 & 171 & 2.256 & sql & 7 & 3.493 \\
doc & 8,025 & 2.266  & text & 188 & 3.176 \\
f & 130 & 3.704      & tmp & 4 & 2.940 \\
fits & 39 & 2.246    & troff & 21 & 3.964 \\ 
gif & 2,026 & 3.551  & ttf & 1 & 2.146 \\
hlp & 7 & 2.805      & txt & 15,008 & 3.496 \\
java & 36 & 3.993    & unk & 434 & 2.864 \\
jpg & 7,167 & 3.793  & wp & 42 & 3.142 \\ 
png & 317 & 3.753    & xls & 4,300 & 2.280 \\
pps & 67 & 2.267     &  &  &  \\
\hline
\end{tabular}
}
\label{tab:header-entropy-excluded}

\end{table}
Many of the file types present in the original data set had a low value calculated for the entropy of their headers. Low entropy files would be trivial to correctly classify using the proposed technique and could bias the experimental results. For this reason, file types in the original data set with a header entropy below 4.0 were excluded from the final analysis. Details of excluded file types with low header entropy are given in Table~\ref{tab:header-entropy-excluded}. The ransomware samples used are all examples of recent high impact strains and are defined in Table ~\ref{tab:ransomwaresamples}. VirusTotal~\cite{VirusTotal} was used to confirm that they were valid ransomware programs.

\begin{table*}[pos=hbpt!,width=7cm,align=\centering]
\begin{center}
\caption {Investigated ransomware samples} \label{tab:ransomwaresamples} 
\begin{tabular}{  l | l  } 
  \hline
  \textbf{Name} & \textbf{SHA256 Hash Value} \\ 
  \hline
  BadRabbit    & 630325cac09ac3fab908f903e3b00d0dadd5fdaa0875ed8496fcbb97a558d0da  \\ 
  NotPetya     & 027cc450ef5f8c5f653329641ec1fed91f694e0d229928963b30f6b0d7d3a745  \\ 
  Phobos       & a91491f45b851a07f91ba5a200967921bf796d38677786de51a4a8fe5ddeafd2  \\ 
  Netwalker    & 57cf4470348e3b5da0fa3152be84a81a5e2ce5d794976387be290f528fa419fd  \\ 
  Ryuk         & d083ecc1195602c45d9cb75a08c395ad7d2b0bf73d7e70e2fc76101c780cc38f  \\
  Sodinokibi   & 96dde0a25cc6ca81a6d3d5025a36827b598d94f0fca6ab0363bfc893706f2e87  \\ 
  Wannacry     & 32f24601153be0885f11d62e0a8a2f0280a2034fc981d8184180c5d3b1b9e8cf  \\ 
  Wastedlocker & 905ea119ad8d3e54cd228c458a1b5681abc1f35df782977a23812ec4efa0288a \\ 
  \hline
\end{tabular}
\end{center}
\end{table*}

Programs used to build data sets containing compressed files were WinZip Version 25, WinRar Version 3.42 and 7Zip Version 9.2.
The resulting test data set contained a total of 84,308 files. 8,685 or 10\%  of the data set were were ransomware encrypted files and the remainder were regular or compressed archive files.

%% file: 06evaluation.tex
\section{Results and Discussion}
\label{cha:evaluation}
\begin{sidewaysfigure*}
  \centering
  \includegraphics[scale=.7]{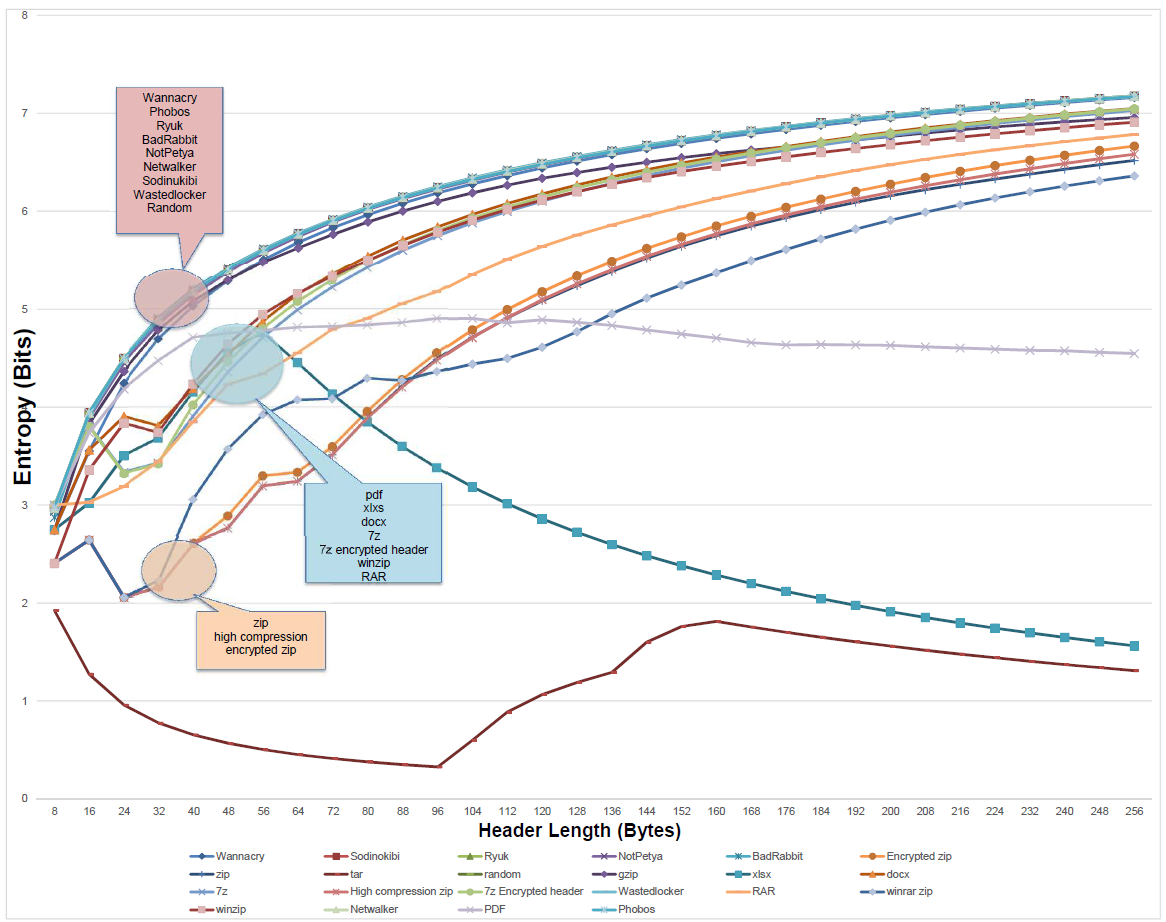}
  \caption{Entropy Values for Different File Types.}
   \label{fig:header_entropy}
\end{sidewaysfigure*}

\noindent Individual entropy values were calculated for each file in the data set. The average entropy was then calculated for each of the file types for header lengths from eight to 256 bytes and the resulting plots for some of these are shown in Figure~\ref{fig:header_entropy}. An example of the data recorded for each file type when examining the first 40 bytes is shown in Table \ref{tab:header-entropy-example}.


\begin{table}
\caption{File entropy values for the first 40 bytes of the file.}
\small
\begin{tabular}{ l | r | c| c}
\hline
\textbf{File type} & \textbf{Sample Size} & \textbf{Entropy} & $\sigma$ \\
\hline
Ryuk & 969 & 5.175 & 0.0785\\
WastedLocker & 979 & 5.170 & 0.0809\\
Sodinokibi & 971 & 5.162 & 0.0808\\
BadRabbit & 556 & 5.111 & 0.0736\\
Netwalker & 3,110 &  5.176 & 0.0798\\
Phobos & 889 & 5.170 & 0.0804\\
gzip & 1,000 & 5.075 & 0.0891\\
NotPetya & 457 & 5.052 & 0.0756\\
Wannacry & 754 & 5.025 & 0.0799\\
PDF & 25,282 & 4.708 & 0.1629\\
KML & 32 & 4.645 & 0.0851\\
PS & 1,212 & 4.616 & 0.1432\\
XML & 932 & 4.605 & 0.2322\\
DWF & 5 & 4.547 & 0.1713\\
EPS & 70 & 4.433 & 0.1582\\
SWF & 43 & 4.303 & 0.5957\\
winzip & 1,000 & 4.232 & 0.1256\\
KMZ & 28 & 4.221 & 0.4118\\
DOCX & 8,038 & 4.190 & 0.0427\\
TEX & 36 & 4.157 & 0.3286\\
XLSX & 4,284 & 4.152 & 0.0437\\
PPTX & 21 & 4.136 & 0.0470\\
RTF & 130 & 4.096 & 0.0944\\
HTML & 25,510 & 4.058 & 0.6840\\
7z Encrypted header & 1,000 & 4.020 & 0.0895\\
7z & 1,000 & 3.903 & 0.0742\\
RAR & 1,000 & 3.847 & 0.1228\\
winrar zip & 1,000 & 3.058 & 0.0118\\
ZIP & 1,000 & 2.612 & 0.0271\\
High compression zip & 1,000 & 2.612 & 0.0268\\
Encrypted zip & 1,000 & 2.611 & 0.0266\\
TAR & 1,000 & 0.654 & 0.0245\\
\hline
\end{tabular}
\label{tab:header-entropy-example}
\end{table}

It can be seen from the entropy graph that as the sample size used for the calculation increases, the entropy values for the compressed files become similar to the entropy values of the encrypted files. However, we believe it is significant that while the initial entropy of the encrypted files is relatively high from the beginning, the files that have been compressed have a much lower entropy value at the start of the files. Other researchers have also noted~\cite{Hall2006, Li2005} that for some file formats, the entropy of the file fragments changes throughout the file. Some explanations for this could be related to the file's magic number, compression techniques being used and in the use of Huffman tables and other metadata at the start of these files contributing to these fragments having a lower entropy than the remainder of the file. 
The only exception to this pattern being files created with the \emph{gzip} program despite this method also using the LZ77 compression algorithm and Huffman tables, similar to the other compression programs tested~\cite{Deutsch1996}. Due to this the dataset for gzip files were not analysed during the classification stage.

The one anomaly to this pattern relates to the values recor-ded for the Phobos strain of ransomware. The relatively low average entropy values for files encrypted with this strain can be explained by the fact that this ransomware sometimes creates files that have large blocks of zeros in the file's header. This has the effect of lowering the average entropy for that file header and the general average for this file type. The entropy calculation algorithm used in the classification was modified to identify these large blocks of zeros and subsequently ignore them when calculating the file header entropy. After implementing this modification the entropy plot for this strain then closely followed the other ransomware samples by also exhibiting higher initial entropy values. 

When examining smaller fragments of the file headers it was observed that there was a measurable average entropy difference between encrypted files having a value of five and compressed files having a value of four.

Previous researchers have commented~\cite{Penrose2013} on the difficulty in discerning the differences in file types of high entropy with little pattern, if any, within the data. Noting that compression algorithms are often optimised for speed, they suggest that it may be possible to further compress already compressed data. As such, encrypted data may be more random than compressed data and, thus, compressibility and randomness may enable these file types to be distinguished from one another. Using the above approach of reducing the sample size used for the analysis of these files and restricting it to the beginning of the file, may enhance the accuracy of identification of the type file being analysed.

An example of the values recorded is given in Table~\ref{tab:header-entropy-example} and illustrates the entropy values of the examined files when analysing the first 40 bytes of the file's header. Differentiation between compressed and encrypted file types is achievable for most ransomware encrypted files as they tend to have an entropy value of at least unity greater than file archives with the exception of the gzip format. The majority of other common file types have significantly lower entropy than compressed files, so these would also be easily distinguishable using this technique.

\subsection{Classification model validation}
\begin{figure*}
    \centering
    \includegraphics[width=1\textwidth]{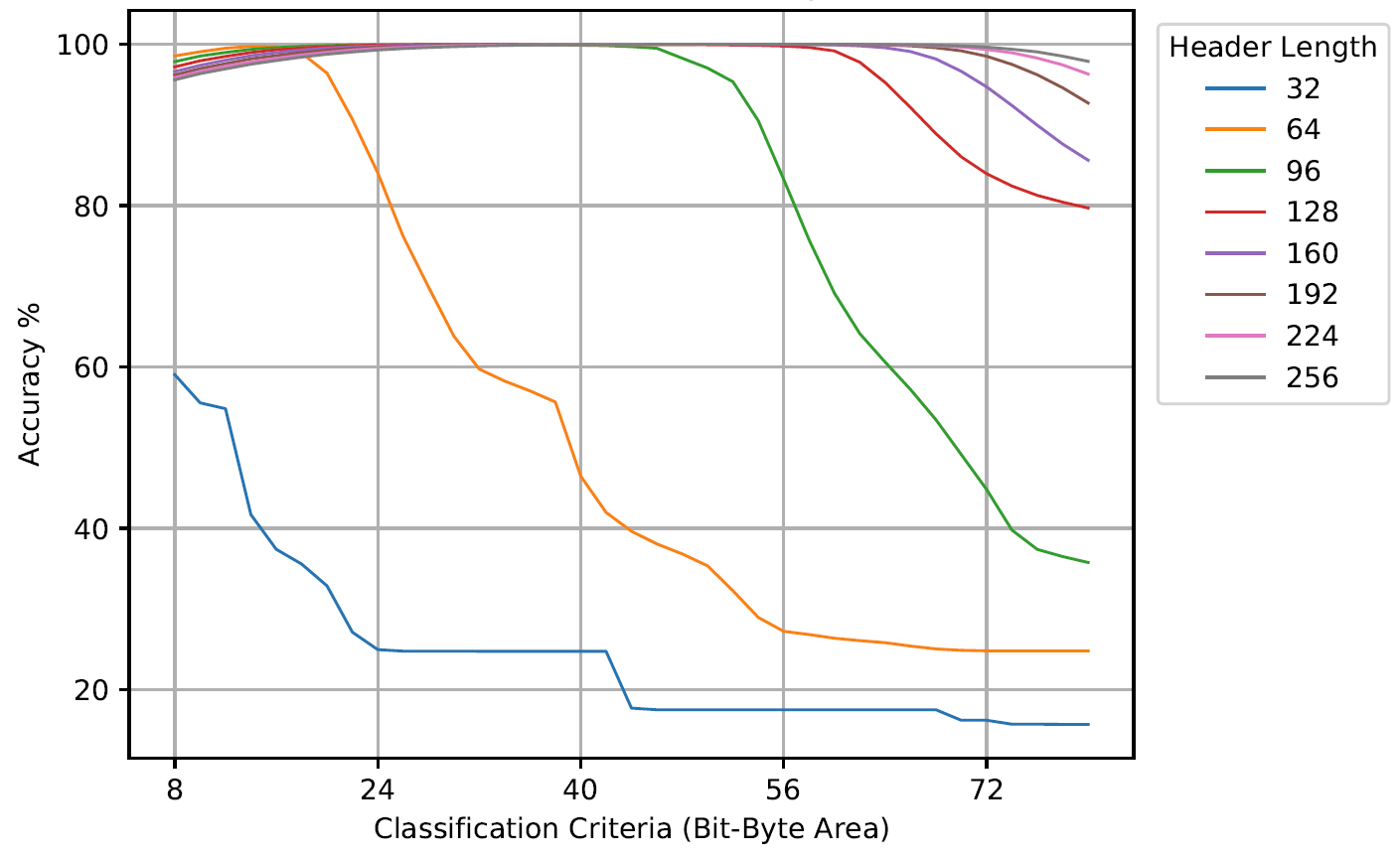}
    \caption{Overall Classification Accuracy.}
    \label{fig:overallclassificationaccuracy}
\end{figure*}

\begin{figure*}
    \centering
    \includegraphics[width=1\textwidth]{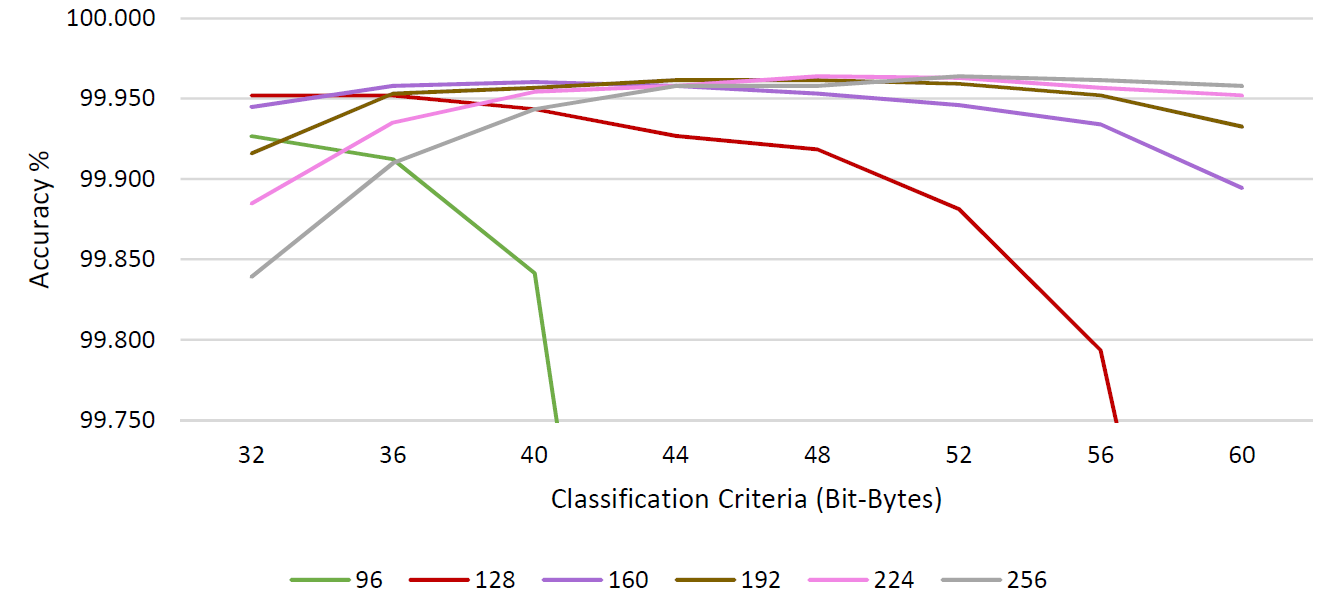}
    \caption{Enlarged section of classification accuracy.}
    \label{fig:zoomedclassificationaccuracy}
\end{figure*}

\noindent The outcome from the second experiment was also successful in identifying encrypted files amongst other files of high entropy and the results are presented in Figure~\ref{fig:overallclassificationaccuracy} with the most interesting portion being highlighted in Figure~\ref{fig:zoomedclassificationaccuracy}. It can be seen that for header lengths between 128 and 256 bytes and, a classification value of between 32 and 56 Bit-Bytes, high rates of success for ransomware encrypted files classification were achieved. Table~\ref{tab:accuracy} shows that in some cases a success classification rate of greater than 99.96\% was achieved with the highest accuracy being highlighted in grey. Related calculated values for this high accuracy region are also provided, with Precision being shown in Table~{\ref{tab:precision}},  Recall in Table~{\ref{tab:recall}} and \emph{F1} in Table~{\ref{tab:f1}}

\begin{table*}
\caption{Classification Accuracy Results (\%)}
\begin{tabular}{ll|lllll}
 \hline
 &  & \multicolumn{5}{c}{Classification Criteria (Bit-Bytes)}  \\
\multirow{7}{*}{\rotatebox[origin=r]{90}{\parbox{3cm} {Header Length \\ 
(Bytes)}}} &  & \textbf{8} & \textbf{24} & \textbf{40} & \textbf{56} & \textbf{72} \\
 \hline
& \textbf{32}  & 63.290 & 20.097 & 19.439 & 14.061 & 11.767  \\
& \textbf{64}  & 99.010 & 87.851 & 46.472 & 24.227 & 19.879  \\
& \textbf{96}  & 98.546 & 99.914 & 99.842 & 87.077 & 51.326  \\
& \textbf{128} & 98.118 & 99.903 & 99.944 & 99.794 & 89.289  \\
& \textbf{160} & 96.736 & 99.825 & \cellcolor[gray]{0.9}99.960 & 99.934 & 96.439  \\
& \textbf{192} & 97.452 & 99.744 & 99.957 & 99.952 & 98.959  \\
& \textbf{224} & 97.234 & 99.631 & 99.954 & 99.957 & 99.517  \\
& \textbf{256} & 97.055 & 99.505 & 99.944 & \cellcolor[gray]{0.9}99.962 & 99.713 \\
 \hline
\end{tabular}
\label{tab:accuracy}
\end{table*}

\begin{table}
\caption{Classification Precision Results (\%)}
\begin{tabular}{ll|llll}
 \hline
 &  & \multicolumn{4}{c}{Classification Criteria (Bit-Bytes)}  \\
\multirow{6}{*}{\rotatebox[origin=r]{90}{\parbox{2.4cm} {Header Length \\ 
(Bytes)}}} &  & \textbf{24} & \textbf{40} & \textbf{56} & \textbf{72} \\
 \hline
 & \textbf{96}  & 99.369 & 98.503 & 44.651 & 17.640 \\
 & \textbf{128} & 100    & 99.462 & 98.058 & 49.324 \\
 & \textbf{160} & 100    & 99.656 & 99.371 & 74.537 \\
 & \textbf{192} & 100    & 100    & 99.553 & 90.923 \\
 & \textbf{224} & 100    & 100    & 99.621 & 95.596 \\
 & \textbf{256} & 100    & 100    &	99.690 & 97.343 \\
 \hline
\end{tabular}
\label{tab:precision}
\end{table}

\begin{table}
\caption{Classification Recall Results (\%)}
\begin{tabular}{ll|llll}
 \hline
 &  & \multicolumn{4}{c}{Classification Criteria (Bit-Bytes)}  \\
\multirow{6}{*}{\rotatebox[origin=r]{90}{\parbox{2.4cm} {Header Length \\ 
(Bytes)}}} &  & \textbf{24} & \textbf{40} & \textbf{56} & \textbf{72} \\
 \hline
& \textbf{96}  & 99.804 & 100    & 100    & 100       \\
& \textbf{128} & 99.286 & 100    & 100    & 100    \\
& \textbf{160} & 98.480 & 99.965 & 100    & 100    \\
& \textbf{192} & 97.640 & 99.850 & 100    & 100    \\
& \textbf{224} & 96.546 & 99.793 & 99.965 & 99.977 \\
& \textbf{256} & 95.337 & 99.632 & 99.942 & 99.977 \\
 \hline
\end{tabular}
\label{tab:recall}
\end{table}

\begin{table}
\caption{Classification F1 Results (\%)}
\begin{tabular}{ll|llll}
 \hline
 &  & \multicolumn{4}{c}{Classification Criteria (Bit-Bytes)}  \\
\multirow{6}{*}{\rotatebox[origin=r]{90}{\parbox{2.4cm} {Header Length \\ 
(Bytes)}}} &  & \textbf{24} & \textbf{40} & \textbf{56} & \textbf{72} \\
 \hline
& \textbf{96}  & 99.586 & 99.246 & 61.736 & 29.990 \\
& \textbf{128} & 99.533 & 99.730 & 99.019 & 66.063 \\
& \textbf{160} & 99.154 & 99.810 & 99.684 & 85.411 \\
& \textbf{192} & 98.760 & 99.793 &	99.770 & 95.246 \\
& \textbf{224} & 98.202 & 99.781 &	99.793 & 97.738 \\
& \textbf{256} & 97.572 & 99.729 &	99.816 & 98.642 \\
\hline
\end{tabular}
\label{tab:f1}
\end{table}

Errors in classification occurred in circumstances where normal files that had a high header entropy were classified as ransomware encrypted (False Positive) and where files resulting from ransomware encryption that had a low header entropy were classified as normal files (False Negative). For example, for a header length of 256 bytes and a classification criteria of 56 Bit-Bytes there were 32 classification errors. Details of which are show in Table~\ref{tab:classificationerrors}.
\begin{table}
\caption {Example of Classification Errors} \label{tab:classificationerrors} 
\begin{tabular}{  l | c| l  } 
  \hline
  \textbf{Classification} & \textbf{Errors} & \textbf{File types} \\ 
  \hline
 False Positives & 27 & PDF(1) PS(14) SWF(12)  \\ 
  False Negatives & 5 & BadRabbit(1) Phobos(2) \\ && Sodinokibi(1) \\&&WasterLocker(1)\\ 
  \hline
\end{tabular}
\end{table}

Finally, a performance comparison test was conducted to determine the difference in execution time  between file header and full file entropy calculations. A sample set of 200 PDF files with sizes varying from 2~KB to 16~MB was used. A performance improvement of three orders of magnitude was achieved by restricting the entropy calculation to the file's header as opposed to calculating the entropy of the entire file.

%% file: 07conclusions.tex
\section{Conclusion}
\label{cha:conclusion}

\noindent The findings from these experiments support the hypothesis that it is possible to classify the type of file being created by only analysing the first few bytes of that file's header. The results from the first experiment demonstrated that, on average, there is a noticeable entropy difference at the beginning of the file between most files and encrypted files generated by ransomware. The paper also introduces the concept of Bit-Byte area and how it can be used as a criteria on which to base file classification decisions. The second experiment highlighting the high success rates achieved when using these criteria. The model was tested against a mixed data set developed by the researchers that, while based on the govdocs1 corpus, also included additional files that more reflect the types of files found on modern systems such as archives. These additional files are included in the data set as they have similar attributes, and were used to determine if the model was able to identify the encrypted files. The results from the experiments confirming that the model was sufficiently robust enough to cope successfully with these more difficult file classifications. In some cases achieving a classification accuracy of greater than 99.96\%. While~\cite{DeGaspari2020,Pont2020} claim that entropy is not a good measure of encryption, these researchers have only considered the file's entire entropy and not a file fragment as we have done in these experiments.


The proposed technique is also ransomware agnostic as no knowledge of the specific ransomware strain is required. The technique is also resilient to the simple  obfuscation techniques~\cite{mcintosh} of altering magic numbers and file extensions. Finally, as only a relatively small portion of the file is being analysed, successful classification is achieved in a fixed amount of time regardless of file size. Consequently reducing the performance impact of such a check and achieving Continella et al.'s goal of minimising the time to decision~\cite{Continella}.

As the specific type of file being created is not the focus of the technique, rather only trying to determine the probability that the file is a result of a ransomware encryption event or not, this technique could possibly be incorporated into a ransomware detection system that intercepts the file writing process. Then, based on the results of the analysis a decision could be made on whether to allow the file write operation or not. If it is determined that the file may be being created by a ransomware process, further analysis of the process creating the file could be performed, possibly using live forensics to discover the encryption keys being used~\cite{Davies2020}.

\subsection{Further research}
\noindent Some further areas of research could be to investigate why files generated using the \emph{gzip} program has a higher initial entropy value than other compression programs using the LZ77 algorithm and Huffman tables. {Investigation into the behaviour of the Phobos ransomware strain may prove beneficial in discovering why it places large blocks of zeros at the beginning of some of the files it creates.

While in these experiments, Shannon entropy was used, it is thought that in the future other calculations maybe be researched. These could include Hamming distance~\cite{Hamming1950}, NIST statistical test suit~\cite{Rukhin2010}, Higuchi fractal dimension~\cite{Kesic2016}, n-grams~\cite{Li2005},  byte frequency analysis~\cite{Roussev2009} or techniques based on machine learning.